\documentclass[aps,prb,twocolumn,showpacs,superscriptaddress]{revtex4-1}
\usepackage{amsfonts}
\usepackage{amsmath}
\usepackage{amssymb}
\usepackage{graphicx}
\usepackage{float}
\usepackage{multirow}
\usepackage{ulem}
\usepackage{gensymb}
\usepackage{dcolumn}
\usepackage[dvipsnames]{xcolor}

\usepackage[bookmarks=false]{hyperref} 
\hypersetup{colorlinks=true,
            linkcolor=blue,
            citecolor=blue,
            urlcolor=orange}
\begin{document}
\title{Exchange interactions in iron and nickel: DFT+DMFT study in paramagnetic phase}
\author{A. A. Katanin}
\affiliation{Center for Photonics and 2D Materials, Moscow Institute of Physics and Technology, Institutsky lane 9, Dolgoprudny, 141700, Moscow Region, Russia}
\affiliation{M. N. Mikheev Institute of Metal Physics of Ural Branch of Russian Academy of Sciences, S. Kovalevskaya Street 18, 620990 Yekaterinburg, Russia}
\author{A. S. Belozerov}
\affiliation{Department of Physics, University of Hamburg, 20355, Hamburg, Germany}
\author{A. I. Lichtenstein}
\affiliation{Department of Physics, University of Hamburg, 20355, Hamburg, Germany}
\author{M. I. Katsnelson}
\affiliation{Institute for Molecules and Materials, Radboud University, Heijendaalseweg 135, 6525AJ Nijmegen, The Netherlands}


\begin{abstract}
We analyze possible ways to calculate magnetic exchange interactions within the density functional theory plus dynamical mean-field theory (DFT+DMFT) approach in the paramagnetic phase. Using the susceptibilities obtained within the ladder DMFT approach {together with} the random phase approximation result for the Heisenberg model, we obtain bilinear exchange interactions. We show that the earlier obtained result of Stepanov {\it et al}. [Phys. Rev. Lett. {\bf 121}, 037204 (2018); {Phys. Rev. B {\bf 105}, 155151 (2022)}] corresponds to {considering individual magnetic moments in each orbital in} the leading-order approximation in the non-local correlations. We {consider a more general approach and apply it} to evaluate the effective magnetic parameters of iron and nickel. {We show that the analysis, based on the inverse orbital-summed susceptibilities, yields reasonable results for both, weak and strong magnets. For iron we find, in the low-temperature limit, the exchange interaction $J_0\simeq 0.20$~eV,
while for nickel we obtain $J_0\simeq 1.2$~eV. The considered method also allows one to describe the spin-wave dispersion at temperatures $T\sim T_C$, which is in agreement with the experimental data}. 
\end{abstract}
\maketitle
\section{Introduction} 
The understanding of magnetic properties turned out to be possible only after the creation of quantum mechanics. In particular, the key role in the physics of magnetic ordering is played by the essentially quantum concept of exchange interactions, 
introduced in the 1920s by Heisenberg, Dirac, Van Vleck, Frenkel, and others (for the historical review as well as for the discussions of basic ideas, we refer to Refs. \onlinecite{vonsovsky,goodenough,mattis,white,yosida}). For magnetic insulators, the Heisenberg Hamiltonian of the localized spins interacting via the exchange interactions is usually considered as the base for the microscopic theory. Itinerant-electron magnets, including elemental iron, cobalt, and nickel, form an important class of magnetic materials \cite{moriya,kubler}. In most of these materials, $d$ electrons responsible for magnetism combine itinerant (band like) and localized (atomic like) features, which makes the problem especially difficult (for a historical review, see Refs.~\onlinecite{vonsovsky,VK1989}). Iron impurities in alkali metals provide a nice example of the evolution from an almost purely atomic regime (Fe in Cs) to an almost purely itinerant one (Fe in Li), with all intermediate steps \cite{alkali}. 

Even for itinerant-electron systems, one can build an effective Heisenberg Hamiltonian and thus determine the exchange parameters by consideration of the variation of the total energy with respect to small deviations in the directions of local magnetic moments ({using magnetic force theorem}) \cite{LKG1984,LKAG1987,KL2000,Kats2004,Solovyev2021,NNNFe1,NNNFe2,gFe}, {rotation of spins by finite angle (consideration of the energies of various spin-density-wave states) \cite{Sandratskii,Gornostyrev}}, {or from the momentum dependence of inverse susceptibilities \cite{Antropov,Bruno,Kats2004,Solovyev2021}}. The corresponding approaches are now broadly used for quantitative calculations of the exchange parameters in specific materials; for a review, see Ref.~\onlinecite{ALMIES3}. However, these approaches require consideration of the symmetry broken state, and the obtained exchange interaction depends on the chosen state (difference between ``local'' and ``global'' spin models \cite{ALMIES3}).

Mostly, these calculations are done based on the density functional theory (DFT) \cite{kubler} which, however,  has difficulties with taking into account atomic-like correlation effects, such as terms and multiplets clearly observed in some systems by photoemission \cite{alkali}. One of the ways to solve this problem and to take into account the mentioned itinerant-localized duality is an idea \cite{anisimov1997,LK1998,KL1999} to combine DFT with the dynamical mean-field theory (DMFT) \cite{DMFT_rev}, which allows one to interpolate between band and atomic limits; for a review of the DFT+DMFT approach, see Ref.~\onlinecite{DFTplusDMFT}. 
The DMFT approach is able to explicitly describe the formation of local magnetic moments due to the Hund interaction (the so-called Hund metals; see Refs.~\onlinecite{Hund1,Hund2,Hund3,Hund4}). The DFT+DMFT calculations were performed for the broad range of magnetic systems, including strong \cite{Licht,OurFe,leonov_fe,OurGamma,Sangiovanni} and weak \cite{OurCr,OurZrZn2,epsilon_fe} itinerant magnets. In most of these systems, fully or partly formed local magnetic moments were obtained. 


The DFT+DMFT approach has been used to calculate exchange interactions in some systems \cite{NNNFe2,kvashnin,locht,solovyev,AKAS1,AKAS2}. In most of the cases, the calculations were performed only for magnetically ordered (that is, symmetry-broken) states. The paramagnetic state was considered \cite{pourovskii,pour} within the ``Hubbard-I'' approximation \cite{LK1998} which is applicable only in the atomic limit, for example, for rare-earth metals \cite{locht}, but definitely not for itinerant-electron three-dimensional magnets. The supercell DMFT theory was used for the calculation of exchange interactions in Refs.~\onlinecite{AKAS1,AKAS2}. However, considerations of the incommensurate correlations require sufficiently large supercells. The systematic account of the non-local effects can be performed in the non-local diagrammatic extensions of DMFT \cite{OurRev}. In particular, the calculation of exchange interactions within the non-local extensions of DMFT (the dual boson approach \cite{dual}) was proposed in Refs.~\onlinecite{ALMIES1,ALMIES2,ALMIES3}. {{Consideration of inverse susceptibilities for the calculation of exchange interactions in the} strong-coupling limit of the DMFT was also suggested in Refs.~\onlinecite{Otsuki,Krien,IgoshevKatanin}}.

Consideration of the exchange interaction in the symmetric phase provides a possibility of its unbiased evaluation, which is not affected by the assumed type of magnetic order. Such a calculation assumes the existence of well-defined local magnetic moments, which, as we note above, exist in a substantial number of strongly correlated substances, and are well described by the DFT+DMFT approach. For itinerant-electron magnets, the corresponding exchange parameters may be {somewhat} larger than those in the ferromagnetic phase, which signals the presence of a short-range order above the Curie temperature \cite{Heine}.  
These exchange parameters can be used to calculate the thermodynamic and dynamical properties of the materials in the paramagnetic phase, via semi-analytical approaches (see, e.g., Ref.~\onlinecite{KataninBelozerov}), Monte Carlo (see, e. g., Refs.~\onlinecite{MC1,MC2}), or atomistic spin dynamics \cite{UppASD} simulations. Therefore, in the present paper, we extend the existing approaches to calculate exchange interactions in non-local extensions of DMFT in the paramagnetic phase and apply the developed approaches to describe the magnetism of iron and nickel.

{The plan of the paper is the following. In Sect. II, we consider the method of calculation of the magnetic exchange interaction from the inverse momentum-dependent susceptibility, expressed in terms of the local particle-hole irreducible interaction vertex. We show that considering individual magnetic moments of each orbital in the leading order in non-local degrees of freedom yields the result of Refs. \onlinecite{ALMIES1,ALMIES2}, while in the general case we obtain more complex expressions. In Sect. III, we describe the details of the DFT+DMFT approach. In Sect. IV, we present the obtained results for the exchange interactions of iron and nickel. Finally, Sect. V presents our conclusions.}  

\section{Methods of calculation of exchange interactions}

{In Hund metals, the Hund exchange leads to the formation of 
local magnetic moment{s (the so-called spin freezing; see, e.g., Refs. \onlinecite{Hund1,Hund4,Hund3,OurFe,OurGamma,
Sangiovanni,epsilon_fe,OurZrZn2,ToschiHund0,ToschiHund1,ToschiHund2,ToschiHund3,MediciHund}), with simultaneous screening of orbital degrees of freedom (the so-called spin-orbital separation \cite{Hund4,SOS}). To describe spin degrees of freedom
in these systems, we introduce orbital-summed on-site spin operators ${\mathbf S}_{i}=\sum_m {\mathbf S}^m_{i}$, where ${\mathbf S}^m_{i}=(1/2)\sum_{\sigma\sigma'}c^+_{im\sigma}\mbox {\boldmath $\sigma $}_{\sigma\sigma'}c_{im\sigma'}$  is the electron spin operator, $c^+_{im\sigma}$ and $c_{im\sigma}$ are the electron creation and destruction operators at site $i$, $d$-orbital $m$, and spin projection $\sigma$, and $\mbox {\boldmath $\sigma $}_{\sigma \sigma'
}$ are the Pauli matrices.}}

We define the exchange interaction by considering the effective Heisenberg model with the Hamiltonian $H=-(1/2)\sum_{{\bf q}} J_{\bf q} {\mathbf S}_{\mathbf q} {\mathbf S}_{-{\mathbf q}}$, {where $\mathbf S_{\mathbf q}$ is the Fourier transform of operators ${\mathbf S}_{i}$}. 
To extract the exchange parameters $J_{\bf q}$, we relate them to the non-local static {longitudinal} susceptibility $\chi_{\mathbf q}=-\langle \langle S^z_{\mathbf q}|S^z_{-{\mathbf q}}\rangle\rangle_{\omega=0}=\sum_{mm^{\prime}}\hat{\chi}_{\bf q}^{m m^{\prime}}$ (the hats stand for matrices with respect to orbital indexes; $\langle \langle ..|..\rangle\rangle_\omega$ is the retarded Green's function). Specifically, we use the random phase approximation (RPA)-like expression for the Heisenberg model\cite{Izyumov} (cf. Refs.~\onlinecite{Antropov,Otsuki,Bruno,IgoshevKatanin,Solovyev2021}), 
\begin{equation}
J_{\mathbf q}=
\chi_{\rm loc}^{-1}-\chi_{\bf q}^{-1},
\label{JqAvDef}
\end{equation}
where 
$\chi_{\rm loc}=-\langle \langle S^z_i|S^z_i\rangle\rangle_{\omega=0}=\sum_{{  m}{  m}'} \hat{\chi}^{{  m}{  m}'}_{\rm loc}$ is the local spin susceptibility. {The consideration of {\it orbital-summed} susceptibilities in Eq. (\ref{JqAvDef}) qualitatively corresponds to calculation of the energy of the spin spiral with the wave vector ${\bf q}$, and is expected to be applicable for both weak and strong magnets.} 

We furthermore use the representation of the orbital-resolved non-local static spin susceptibility {in the interacting itinerant electrons model. For simplicity, we consider the density-density interaction,
\begin{equation}
{H}_{\rm Coul} = \frac{1}{2}\sum\limits_{i,mm',\sigma\sigma'} U^{mm^\prime}_{\sigma\sigma^\prime}
{n}_{im\sigma} {n}_{im^\prime\sigma^\prime} \label{Hint}
\end{equation}
(${n}_{im\sigma}=c^+_{im\sigma}c_{im\sigma}$ is the particle number operator), for which we obtain} (cf. Refs.~\onlinecite{MyGamma,MyEDMFT,OurRev,EdwHrtz})
\begin{equation}
\hat{\chi}^{}_{\bf q}=\frac{1}{2}\left[\hat{\Pi}_{\bf q}^{-1}-\hat{U}^{s}\right]^{-1}, \label{phi}
\end{equation}
where the static spin polarization {(irreducible static spin susceptibility)}  $\hat{\Pi}_{\bf q}^{}=T\sum_\nu \hat{\chi}_{{\bf q},\nu} ^{0}\hat{\gamma}_{{\bf q},\nu}^{}$,  $\hat{U}^s=\hat{U}_{\uparrow\downarrow}-\hat{U}_{\uparrow\uparrow}$ is the electron interaction matrix in the spin channel, $\nu$ are the fermionic Matsubara frequencies, and the matrix inversions in Eq. (\ref{phi}) are assumed.

The particle-hole irreducible triangular vertex $\gamma$ in the spin channel is 
given by \cite{MyGamma,MyEDMFT,OurRev}
\begin{equation}
    {\hat \gamma}_{{\bf q},\nu}^{{m} {m}'}=\sum\limits_{\nu'} \left[\hat{E}-(\hat{\Phi}_{\nu\nu'}^{}-\hat{U}^s) \hat{\chi} ^{0}_{{\bf q},\nu'}\right]_{\nu {m}, \nu' {m}'} ^{-1},\label{gamma}
\end{equation}
where $\hat{E}$ is the identity matrix with respect to orbital indexes and fermionic frequencies. 
The electron particle-hole irreducible vertex in the spin channel $\hat{\Phi}_{\nu\nu'}$ is assumed to be local (which also yields the dependence of the vertex (\ref{gamma}) on the momentum ${\bf q}$ and frequency $\nu$ only). In the following, we extract the vertex $\hat{\Phi}_{\nu\nu'}$ from the dynamical mean-field theory (see Sect. III). 
The bare susceptibility reads
\begin{equation}
{\hat \chi}^{0,mm'}_{ {\bf q},\nu}=-T\sum\limits_{\mathbf{k}}G^{mm'}_{{\bf k},\nu}G^{m'm}_{{\bf k+q},\nu},\label{chi0}
\end{equation}
where 
$G^{mm'}_{{\bf k},\nu}$ is the electron non-local Green's function.
The local susceptibility can be represented similarly to Eqs. (\ref{phi})-(\ref{chi0}) with the respective local spin polarization $\Pi_{\rm loc}=T\sum_\nu \hat{\chi}_{{\rm loc},\nu} ^{0}\hat{\gamma}_{{\rm loc},\nu}^{}$, local vertex $\gamma_{{\rm loc},\nu}$ (see Appendix A) and ${\hat \chi}_{{\rm loc},\nu}^{0,mm'}=-T (G^{m}_{{\rm loc}, \nu})^2  \delta_{mm'}$, with $G^{m}_{{\rm loc}, \nu}$ being the electron local  Green's function. 

{Let us compare the presented method of evaluation of exchange interaction to that considering magnetic exchange among individual magnetic moments of each orbital} $H=-(1/2)\sum_{{\bf q},{m},{m}'} J^{{ m}{ m}'}_{\bf q} {\mathbf S}^{m}_{\mathbf q} {\mathbf S}^{{m}'}_{-{\mathbf q}}$, where {$\mathbf S^m_{\mathbf q}$ is the Fourier transform of operators ${\mathbf S}^m_{i}$}. The exchange parameters (which are matrices in orbital indexes) in this case can be extracted from
$\hat{J}%
_{\bf q}=  \hat{\chi}_{\text{loc}}^{-1}-\hat{\chi}_{{\bf q}}^{-1}$ 
where the matrix inversions are assumed. Taking into account Eq. (\ref{phi}), the resulting orbital-resolved exchange interaction of the effective Heisenberg model reads 
\begin{equation}
    \hat{J}_{\bf q} =2(\hat{\Pi}_{\rm loc}^{-1}-\hat{\Pi}_{\bf q}^{-1}).
    \label{ExMy}
\end{equation}
To the leading order in the non-local degrees of freedom, we obtain in this case (see Appendix B, cf. Ref.~\onlinecite{IgoshevKatanin})
\begin{equation}
    \hat{J}_{\bf q}=2\hat{\Pi}_{\rm loc}^{-1}\left[\sum_\nu \hat{\gamma}^T_{{\rm loc},\nu}(\hat{\chi}^0_{{\bf q},\nu}-\hat{\chi}^0_{{\rm loc},\nu})\hat{\gamma}_{{\rm loc},\nu}\right]\hat{\Pi}_{\rm loc}^{-1},
    \label{ExALMIES}
\end{equation}
which, as we show in Appendix B, agrees with the result of Refs.~\onlinecite{ALMIES1,ALMIES2} (see also Ref. ~\onlinecite{ALMIES3}).
{{We note that 
{{similarly to the previous studies of} the ordered state}, }}
{Eq. (\ref{ExALMIES}) describes the Ruderman–Kittel–Kasuya–Yosida (RKKY) mechanism of magnetic exchange}.
{In the ferromagnetic state, this} corresponds to extracting exchange interaction
from the poles of dynamical susceptibility~\cite{Kats2004,ALMIES3}, whereas the exchange parameters found from the inverse static susceptibility also contain  nontrivial corrections due to the residue in the pole, related to the non-RKKY contributions. We expect the same {holds for exchange parameters (\ref{ExALMIES}) continued to the {para}magnetic phase,
especially for strong magnets, such as iron (see below). In the ferromagnetic ground state, the above-mentioned corrections vanish in the limit $q \rightarrow 0$, but this may be different for the susceptibility extrapolated from the paramagnetic phase}. 

Averaging the obtained exchange interactions over orbitals with the local susceptibilities, which are proportional to the products of local magnetic moments, one can get the orbital-averaged exchange,  
\begin{equation}
    J_{\bf q}=\frac{1}{\chi_{\rm loc}}\sum_{{  m}{  m}'}( \hat{J}^{{  m}{  m}'}_{\bf q} \hat{\chi}_{\rm loc}^{{  m}'{  m}}).\label{JqAv}
\end{equation}
 {As discussed above, this average assumes that the local magnetic moments on different orbitals exist independently and possess approximately equal magnetic exchange}.  {This approach yields the same result as Eq. (\ref{JqAvDef}) only for the exchange interactions, which are independent on the orbital indices, ${\hat J}_{\mathbf q}=J_{\mathbf q}$}. 
One can also show that the magnetic exchange, extracted from the orbital-summed susceptibilities (\ref{JqAvDef}),
is equivalent to the more sophisticated average of the orbital-resolved exchange interactions of Eq. (\ref{ExMy}),%
\begin{equation}
J_{\bf q}=\frac{\sum_{mm^{\prime}} \left[  \hat{\chi}_{\text{loc}}\hat
{J}_{\bf q}\hat{\chi}_{\bf q}\right]  _{m m^{\prime}}}%
{\sum_{mm^{\prime}}\hat{\chi}_{\text{loc}}^{m m^{\prime}}\sum_{mm^{\prime}}\hat{\chi}_{\bf q}^{m m^{\prime}}}.%
\end{equation}

{We note that the density-density interaction (\ref{Hint}) assumes Ising symmetry of the Hund exchange and therefore violates the SU(2) symmetry, and is used to simplify calculations of the vertices. We restore the SU(2) symmetry of the magnetic exchange by assuming that all components of magnetic exchange are equal to the longitudinal interaction, co-aligned with the direction of Hund exchange, and determined by the Eq. (\ref{JqAvDef}). At the same time, we emphasize that the formalism of this section is applicable to the rotationally invariant SU(2) interaction as well by considering the matrices of the size $n_{\rm orb}^2\times n_{\rm orb}^2$ instead of $n_{\rm orb}\times n_{\rm orb}$, where $n_{\rm orb}=5$ is the number of correlated $d$ orbitals.}


\section{DFT+DMFT calculation}

For application of the described method we below consider $\alpha$, $\gamma$-iron, and nickel. Our DFT calculations were performed using the full-potential linearized augmented-plane wave method implemented in the ELK code\cite{ELK} supplemented by the Wannier function projection procedure (Exciting-plus code\cite{Plus}).
%
%
The Perdew-Burke-Ernzerhof form of generalized gradient approximation was used.
For a bcc Fe we adopt the experimental lattice constant of 2.91 \AA~(Ref.~\onlinecite{bcc_fcc_iron_alat}), for fcc iron, 3.65~\AA~  (Ref.~\onlinecite{bcc_fcc_iron_alat}), and for nickel 3.52~\AA~ 
 (Ref.~\onlinecite{nickel_alat}).
The convergence threshold for total energy was set to $10^{-8}$~Ry.
The reciprocal space integration was performed using ${18\times 18\times 18}$\, ${\bf k}$-point grid.

To take into account the on-site Coulomb correlations, we consider the following Hamiltonian within the DMFT approach:
\begin{equation}
{H}_{\rm DMFT} = {H}_{\rm DFT}^{\rm WF} + {H}_{\rm Coul} - {H}_{\rm DC},
\end{equation}
where ${H}_{\rm DFT}^{\rm WF}$ is
the effective low-energy Hamiltonian constructed as a projection of Kohn-Sham states obtained within DFT onto the basis of Wannier functions, 
${H}_{\rm Coul}$ is the local Coulomb interaction Hamiltonian,
and ${H}_{\rm DC}$ is the double-counting correction for accounting of the on-site interaction already described by DFT.

For construction of ${H}_{\rm DFT}^{\rm WF}=\sum_{{\bf k},\lambda\lambda',\sigma}H_{\bf k}^{\lambda\lambda'} c^+_{{\bf k}\lambda\sigma} c_{{\bf k}\lambda'\sigma}^{}$, we use a basis set of atomic-centered Wannier functions~\cite{Wannier}, $\lambda$ is the orbital index, corresponding to the 
considered $3d$, $4s$ and $4p$ states, and $c^+_{{\bf k}\lambda\sigma}$ and $c_{{\bf k}\lambda\sigma}$ are the corresponding creation and annihilation operators. We consider the local Coulomb interaction in the density-density form (\ref{Hint}).
For parametrization of the local interaction matrix $U^{mm^\prime}_{\sigma\sigma^\prime}=U^{mm^\prime}-I^{mm'}\delta_{\sigma\sigma'}$
we use Slater integrals $F^0$, $F^2$, and $F^4$ linked to the Hubbard parameter ${U\equiv F^0}$ and Hund's rule coupling ${J_{\rm H}\equiv (F^2+F^4)/14}$, such that (see Ref.~\onlinecite{u_and_j})
\begin{align}
U^{mm'}&=\sum_{k=0,2,4} a_k({m,m',m,m'}) F^k,\notag \\
I^{mm'}&=\sum_{k=0,2,4} a_k({m,m',m',m}) F^k,
\end{align}
where 
\begin{align}
a_k({m,m',m'',m'''})&=\frac{4\pi}{2k+1}\sum_{q=-k}^k \langle m | Y_{kq} | m''\rangle \notag \\&\times  \langle m'| Y^*_{kq} |m'''\rangle,
\end{align}
$| m\rangle$ is the $d$ state with magnetic number $m$, and $Y_{kq}$ is the spherical harmonic.
%
We perform our calculations with ${U=4}$~eV and ${J_{\rm H}=0.9}$~eV, which were used in previous DFT+DMFT studies of exchanges interactions in iron\cite{AKAS1,AKAS2}. We use a double-counting correction ${H}_{\rm DC} = \sum_{im\sigma} M^m_{\rm DC} {n}_{im\sigma} $ in the around mean-field form~\cite{AMF},  $M^m_{\rm DC}=\langle {n}_{id} \rangle [U (2 n_{\rm orb} {-} 1) - J_{\rm H}  (n_{\rm orb} {-} 1)] / (2 n_{\rm orb})$, where
${n}_{id}$ is the {operator of the} number of $d$ electrons at the site $i$.
%
%

In DMFT, the lattice problem is mapped onto an effective impurity problem, 
\begin{align}
S_{\rm imp} &=- \int_0^\beta d\tau \int_0^\beta d\tau^\prime \sum_{m\sigma} \mathfrak{c}_{m\sigma}^\dag(\tau)\mathcal{G}_{0}^m(\tau-\tau^\prime)^{-1}\, \mathfrak{c}_{m\sigma}(\tau^\prime) \notag\\&+ 
\frac{1}{2}\int_0^{\beta} d\tau\sum\limits_{mm',\sigma\sigma'}\, 
U_{mm^\prime}^{\sigma\sigma^\prime}
\mathfrak{n}_{m\sigma}(\tau) \mathfrak{n}_{m^\prime\sigma^\prime}(\tau),
\end{align}
where $\tau$ and $\tau^\prime$ are imaginary times, 
$\mathfrak{c}_{m\sigma}^\dag$ ($\mathfrak{c}_{m\sigma}$)
are the Grassmann variables for the creation (annihilation) of an electron with spin~$\sigma$ at orbital~$m$, and $\mathfrak{n}_{m\sigma}=\mathfrak{c}^+_{m\sigma}\mathfrak{c}_{m\sigma}$,
$\mathcal{G}_{0}^m(\tau-\tau^\prime)$ is the bath Green's function.
%
To solve this quantum impurity problem, we employ the hybridization expansion continuous-time quantum Monte Carlo method~\cite{CT-QMC}, realized in the iQIST package \cite{iQIST}.
%

{From the obtained electron self-energy at the impurity site $\Sigma^m_\nu$ we obtain the non-local Green's functions,
\begin{equation}
G^{mm'}_{{\bf k},\nu}=\left[i\nu+\mu-H_{\bf k}^{\lambda \lambda'}-(\Sigma^\lambda_\nu -M_{\rm DC}^\lambda)\delta_{\lambda\lambda'}\right]^{-1}_{mm'},\label{Gk}
\end{equation}
where $\mu$ is the chemical potential, determined by fixing the number of valent electrons, and we put both the self-energy $\Sigma^\lambda_{\nu}$ and the double counting term $M^\lambda_{\rm DC}$ to zero for the uncorrelated $s,p$ orbitals. The Fourier transform of the bath Green's function $\mathcal{G}_{0,\nu}^m$ is obtained from the self-consistent condition equating the local Green's function $G_{{\rm loc},\nu}^m=((\mathcal{G}_{0,\nu}^m )^{-1}-\Sigma^m_\nu)^{-1}$ to the local part of the lattice Green's function (\ref{Gk}).}

\begin{figure}[t]
		\center{
				\vspace{0.3cm}
		\includegraphics[width=0.9\linewidth]{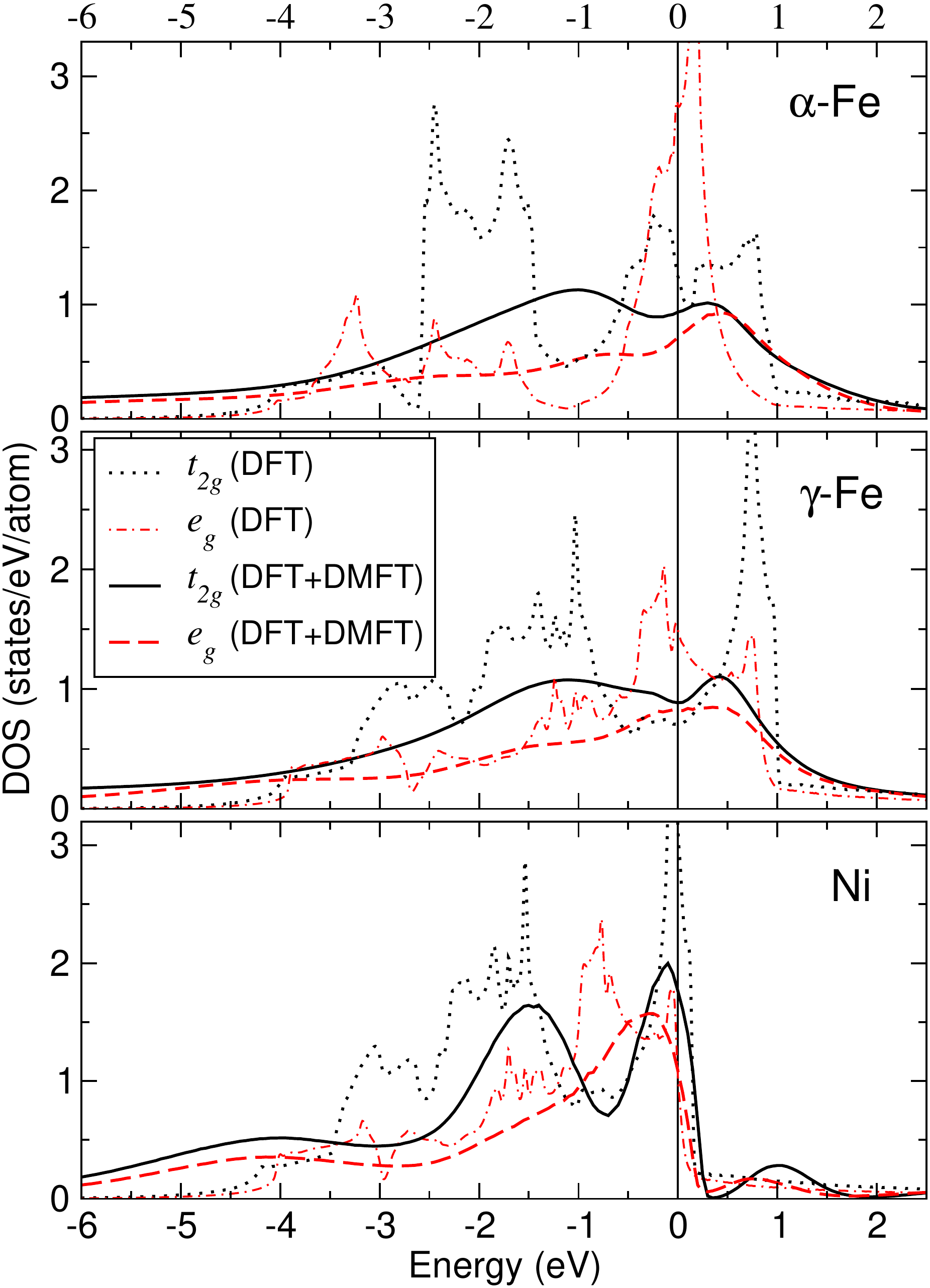}}
		\caption{
		Orbital-resolved DFT (dotted and dash-dotted lines) and DFT+DMFT (solid and dashed lines) density of states of $\alpha$-iron (top panel), $\gamma$-iron (middle panel) and nickel (bottom panel). DFT+DMFT densities of states are calculated at the inverse temperature $\beta=5$~eV$^{-1}$ for $\alpha$-iron and $\beta=10$~eV$^{-1}$ for $\gamma$-iron and nickel. The Fermi level corresponds to zero energy.}
\label{FigDOS}
\end{figure}


We present the orbital-resolved densities of states in the DFT and DFT+DMFT approaches in Fig. \ref{FigDOS}; see also Refs. \onlinecite{OurFe,OurGamma}. One can see that in $\alpha$-iron, the DFT+DMFT density of states is strongly suppressed at the Fermi level, {which is the signature of the formation of local magnetic moments \cite{OurFe}}. For $\gamma$-iron and nickel we find weaker suppression, which characterises these substances as {having partly formed local magnetic moments \cite{OurGamma,Sangiovanni}}. The formation of local magnetic moments in these substances is governed by Hund exchange \cite{OurFe,OurGamma,Sangiovanni}. {In nickel, similarly to the previous calculation in the ferromagnetic state \cite{Licht,LichtNi}, we observe the lower- and upper Hubbard bands at the energies $\sim-4$~eV and $1$~eV, which originate from the narrower bandwidth of nickel in comparison to $\alpha$-iron and $\gamma$-iron}. 

{{After the DMFT self-consistent procedure is performed, the local dynamic spin vertex  $F^{mm'}_{\nu\nu'}$ is evaluated from the continuous time quantum Monte-Carlo (CT-QMC) approach 
(the pairs of the incoming and outgoing particles are characterized by the orbital indexes $m,m'$ and frequencies $\nu,\nu'$). The particle-hole irreducible vertex ${\hat \Phi}_{\nu \nu'}$} is evaluated via the Bethe-Salpeter equation,
\begin{align}
(\hat F)^{-1}_{\nu m,\nu' m'}&=(\hat \Phi)_{\nu m,\nu' m'}^{-1}
-\chi_{{\rm loc},\nu}^{0,mm'} \delta_{\nu\nu^{\prime}}.
\label{local_BS}%
\end{align}
In the calculation of vertices we account for $60$-$90$ fermionic frequencies (both positive and negative). In the summations over frequencies, the corrections on the finite size of the frequency box are accounted according to Refs.~\onlinecite{My_BS,MyEDMFT}; see also Appendix A. 

We use Eq. (\ref{JqAvDef}), as well as {Eqs. (\ref{ExMy}) and (\ref{ExALMIES}) together with Eq. (\ref{JqAv})}, to calculate exchange interactions for the converged DFT+DMFT solution. We furthermore subtract the small local part of the resulting exchange interaction (the local part of Eq. (\ref{ExALMIES}) vanishes identically; for Eq. (\ref{ExMy}) it constitutes approximately 10\% of the exchange interaction, while for Eq. (\ref{JqAvDef}) it constitutes about 3\% of the exchange interaction).

\section{Results for the exchange interactions}

\begin{figure}[b]
		\center{
		\includegraphics[width=0.9\linewidth]{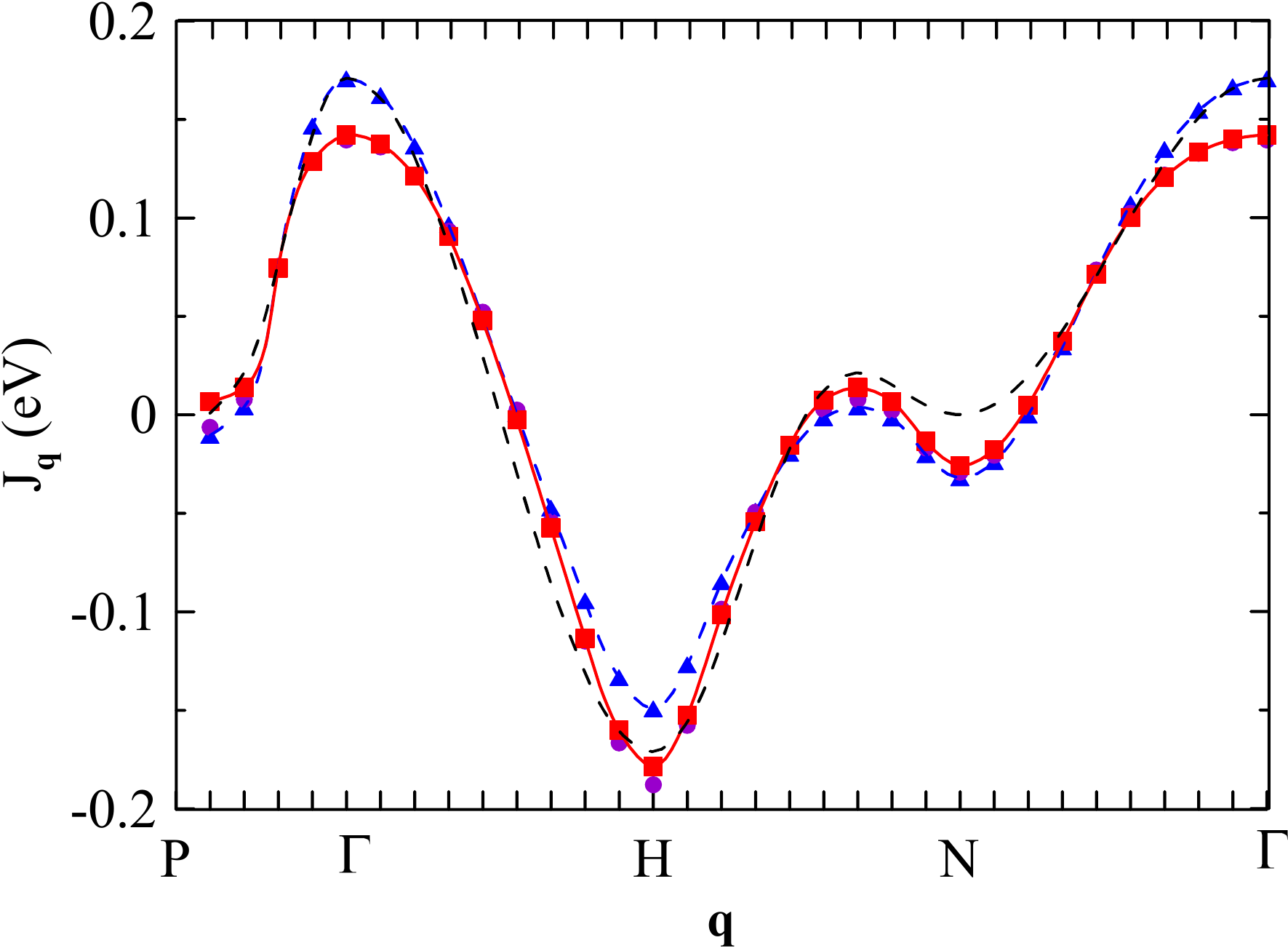}}
		\caption{
		Momentum dependence of exchange interactions in $\alpha$-Fe at $\beta=5$~eV$^{-1}$ along the symmetric directions. The violet dot-dashed line (dots) corresponds to the Eq. (\ref{ExMy}) with the orbital average (\ref{JqAv}); the blue dashed line (triangles) corresponds to the leading-order approximation (\ref{ExALMIES}) with the average (\ref{JqAv}); and the red solid line (squares) corresponds to the to the result from the orbital-summed susceptibilities, given by Eq. (\ref{JqAvDef}). The dashed line without symbols shows the momentum dependence of the nearest-neighbor exchange.}
		 \label{FigJq_aFe}
\end{figure}

We consider first the results for the $\alpha$ (bcc) phase of iron. Figure~{\ref{FigJq_aFe}} shows the momentum dependence of the obtained exchange interaction at $\beta=5$~eV$^{-1}$ near the DMFT Curie temperature $T_{\rm C}=2130$~K. All considered approaches give close values of the exchange interaction, which is related to the presence of well-defined local magnetic moments in $\alpha$-Fe; the magnetic exchange determined from the Eq.~(\ref{ExALMIES}) is somewhat larger than in the other considered approaches. {The obtained momentum dependencies of the exchange interaction somewhat deviate from the
nearest-neighbor exchange interaction, showing the important role of the long-range interactions; cf. Refs.~\onlinecite{NNNFe1,NNNFe2,kvashnin,NNNFe3} and references therein}.

  \begin{figure}[t]
		\center{		\includegraphics[width=0.95\linewidth]{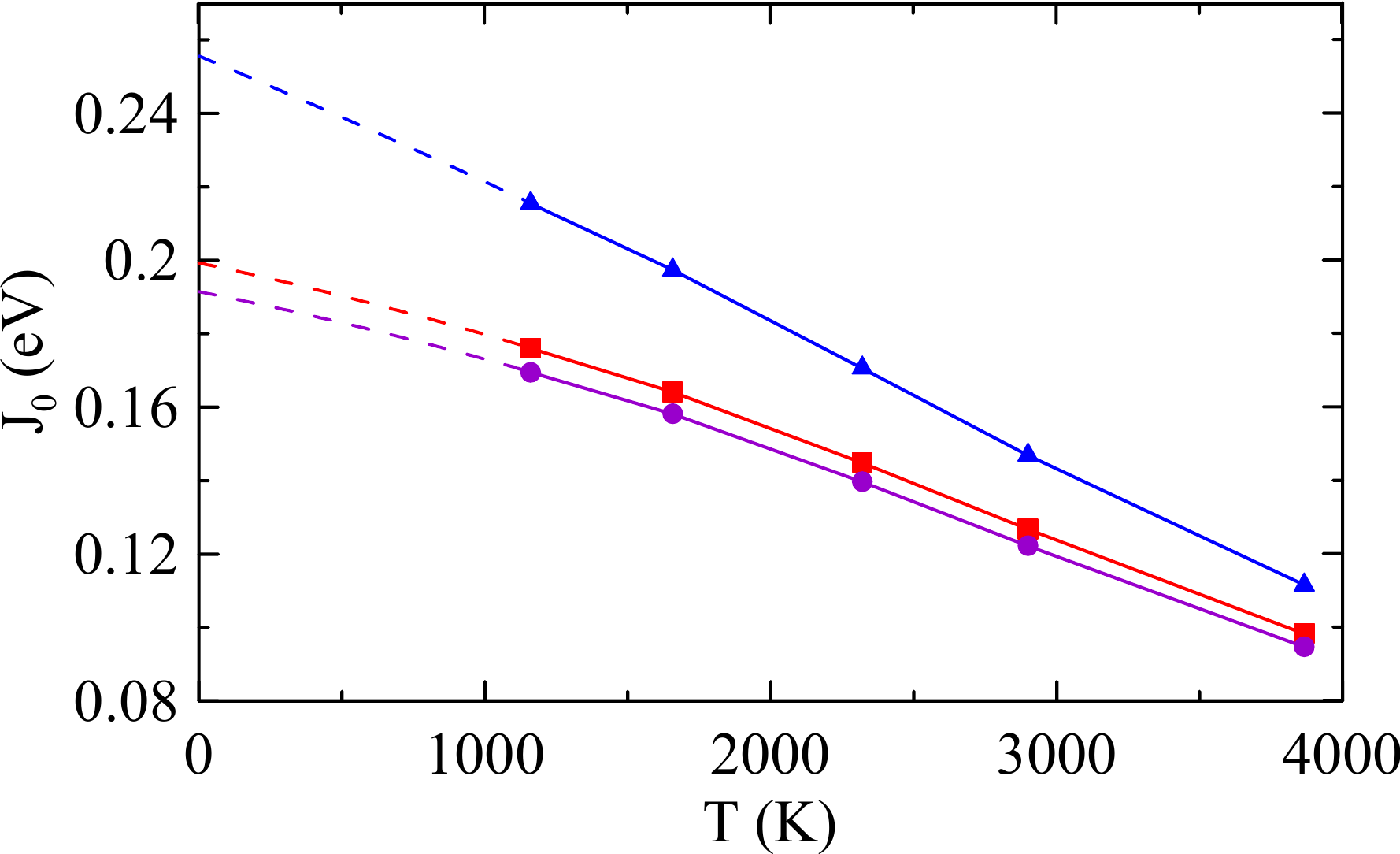}}
		\caption{Temperature dependence of the exchange interactions in $\alpha$-Fe (solid lines). The notations are the same as in Fig. \ref{FigJq_aFe}. The extrapolation to the zero-temperature limit is shown by dashed lines.}
\label{FigJT_aFe}
\end{figure}

In contrast to the DFT (supercell DMFT) approach, which is applicable at $T=0$ ($T>T_C$), the present method gives a possibility to study the temperature dependence of the magnetic interaction in a broad temperature range. The temperature dependence of the exchange interaction is shown in Fig.~\ref{FigJT_aFe}; see also the Supplemental Material \cite{SM} for more details. One can see that the results of the leading-order approximation, given by Eq. (\ref{ExALMIES}), show a stronger increase with decrease of temperature.
The low-temperature exchange integral in iron, extracted from the orbital-summed susceptibilities, given by Eq.  (\ref{JqAvDef}), is estimated as $J_0\simeq 0.2$~eV, which is close to the value previously obtained within the supercell DMFT approach \cite{AKAS1} and the local spin density functional approach (using the local magnetic force theorem) \cite{LKAG1987,Kats2004}. The latter approach yields \cite{Kats2004} $J^{\rm DFT}_0 \mu_{\rm DFT}^2/(g\mu_B)^2=0.27$~eV ($g=2$ is the $g$ factor, $\mu_B$ is the Bohr magneton), which using the DFT magnetic moment $\mu_{\rm DFT}=2.2\mu_B$ \cite{kvashnin} implies $J^{\rm DFT}_0=0.22$~eV.
{At the same time, from the lowest-order non-local correction, given by Eq.~(\ref{ExALMIES}), we obtain a somewhat larger exchange interaction $J_0\simeq 0.26$~eV at low temperatures.}

{As we discuss in Supplemental Material \cite{SM}, the obtained exchange interactions are relatively weakly dependent on the reasonable choice of the Coulomb interactions $U$ and $J_H$, even though the self-energies more strongly depend on these parameters. This is due to the compensation of the self-energy and vertex corrections, contained in $\hat \gamma_{{\bf q},\nu}$. The latter is different from the standard DFT+DMFT approach to the calculation of exchange interactions in the ordered phase (see, e.g., Refs. \onlinecite{KL2000,kvashnin,ALMIES3}), which relies on an approximate form of the vertex corrections (see, e.g., Ref. \onlinecite{ALMIES3}), which may affect exchange interactions, and, in particular, lead to the stronger dependence dependence of $J_{\mathbf q}$ on the parameters of the Coulomb interaction \cite{kvashnin,NNNFe2}.}

  \begin{figure}[t]
		\center{		\includegraphics[width=0.8\linewidth]{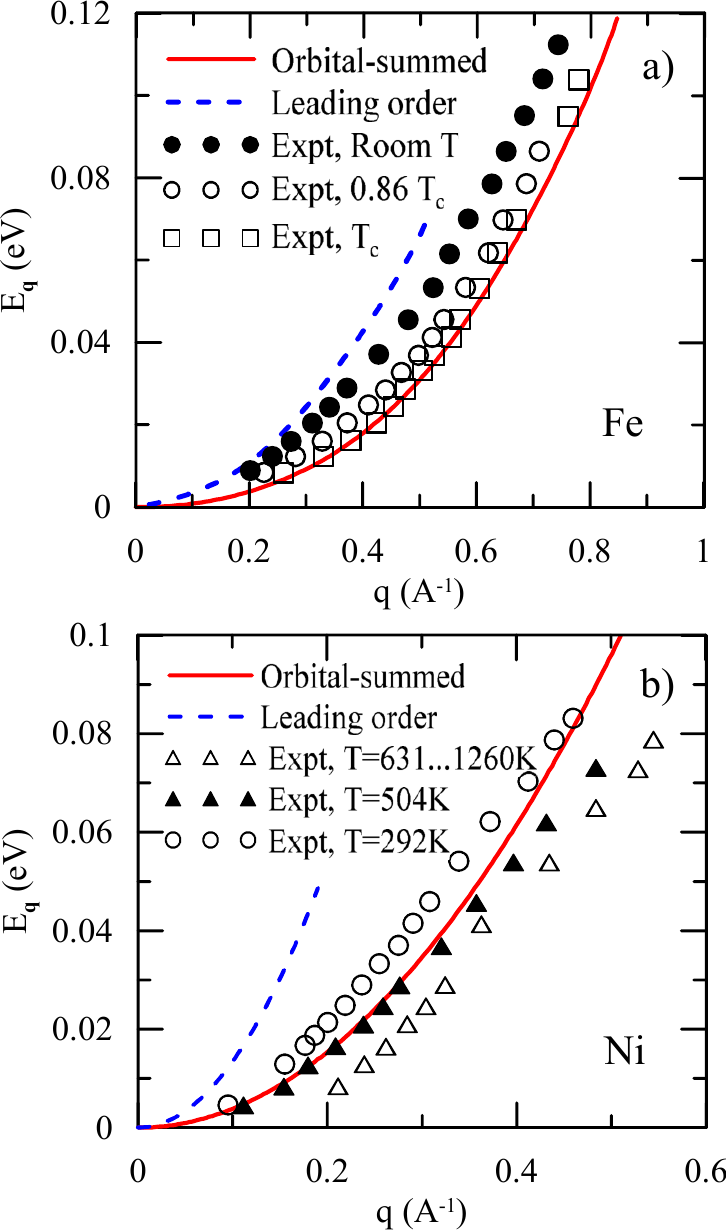}}
		\caption{Comparison of the experimental magnon dispersion in (a) $\alpha$-Fe along the [110] direction (Ref. \onlinecite{Lynn1}) and (b) nickel along the [111] direction (Ref. \onlinecite{Lynn2}) at various temperatures (symbols) with the results of DFT+DMFT approach at $\beta=10$~eV$^{-1}$. The solid red line corresponds to the orbital-summed exchange interactions (\ref{JqAvDef}); the dashed blue line is the result of the leading-order non-local approximation of Eqs. (\ref{ExALMIES}) and (\ref{JqAv}).  }
\label{FigD}
\end{figure}

{In Fig. \ref{FigD}(a), we show the comparison of the experimental magnon dispersions to those obtained as $E_{\mathbf q}=p_{\rm loc}(J_0-J_{\mathbf q})$, where the effective spin $p_{\rm loc}$ is determined by $p_{\rm loc}(p_{\rm loc}+1)=(\mu_{\rm loc}/g\mu_B)^2\equiv 3/(d \chi^{-1}_{\rm loc}/dT)$; $p_{\rm loc}\simeq 1.5$ for $\alpha$-iron. One can see that the dispersion, obtained from the orbital-summed susceptibilities (which is also close, for $\alpha$-iron, to that from the orbital-resolved inverse susceptibilities (\ref{ExMy}) with the average (\ref{JqAv})) agrees well with the experimental data. At the same time, the dispersion, obtained from the leading-order approximation (\ref{ExALMIES}) and (\ref{JqAv}), somewhat overestimates the spin-wave stiffness.} 


  \begin{figure}[t]
		\center{
\includegraphics[width=0.95\linewidth]{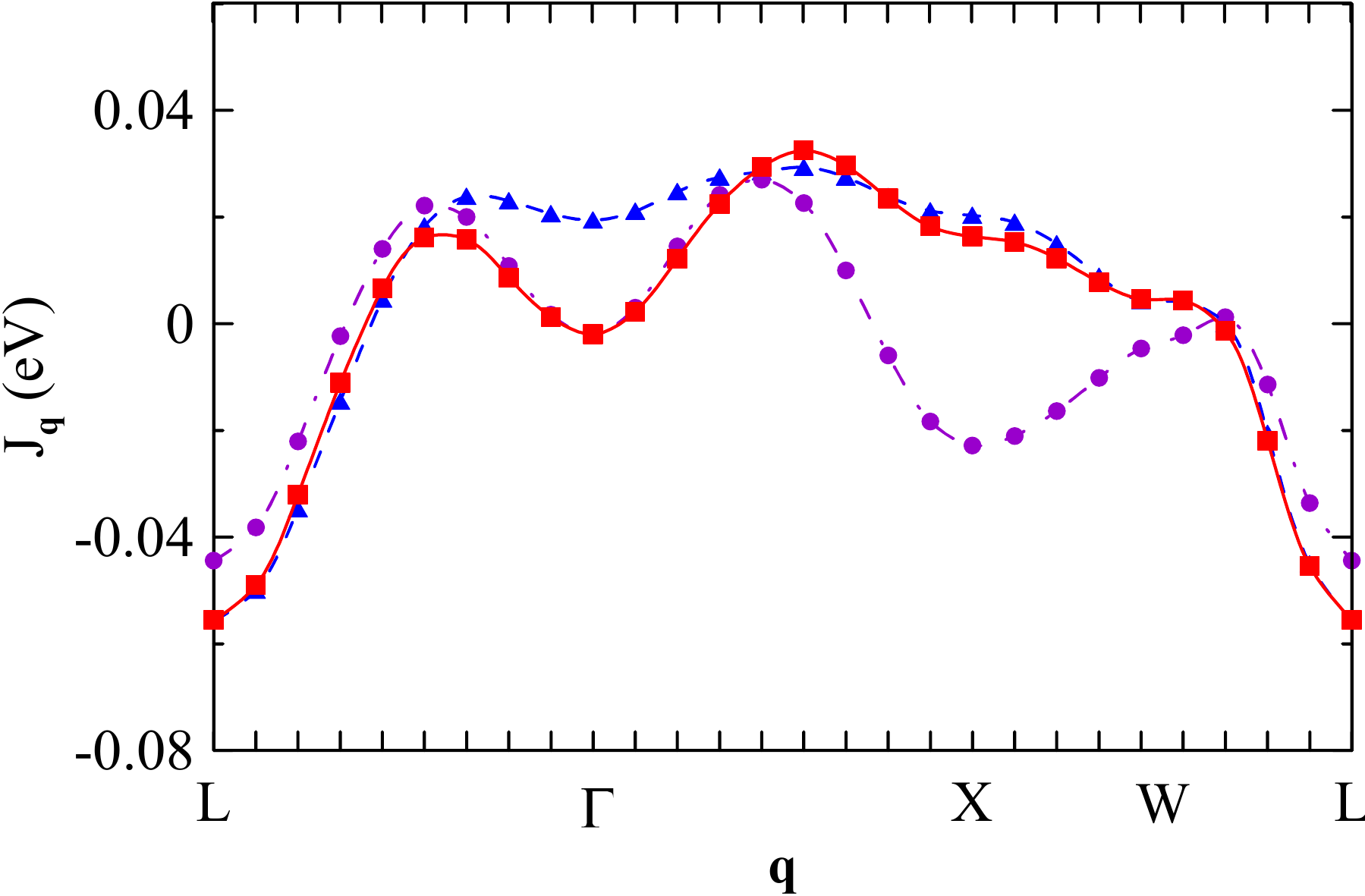}}
		\caption{
		Momentum dependence of exchange interaction in $\gamma$-Fe at $\beta=10$~eV$^{-1}$ along the symmetric directions. The notations are the same as in Fig. \ref{FigJq_aFe}}.
\label{FigJq_gFe}
\end{figure}

By construction, the  method of calculation of the exchange interaction, which uses orbital-summed suscepibilities in Eq.~(\ref{JqAvDef}), allows us to 
reproduce the DMFT Curie temperature from the mean-field-like equation $T_C=J_0(T_C) \mu_{\rm loc}^2(T_C)/(3g^2\mu_B^2)$ (we find $J_0(T_C)\simeq0.15$~eV, $\mu^2_{\rm loc}(T_C)=14.7\mu_B^2$). Although the Curie temperature in DMFT is overestimated almost two times in comparison to the experimental data, we note that the obtained square of the local magnetic moment is also overestimated with respect to the experimental data and the local magnetic moment from the uniform susceptibility $\mu^2=9.8 \mu_B^2$. Therefore, we can relate the corrections to the above-cited mean-field result (or, equivalently, RPA formula (\ref{JqAvDef})) {at least partly} to the overestimate of the local magnetic moment in DMFT. The remaining part of the overestimation of the Curie temperature, which may  also yield somewhat smaller exchange integrals and/or local magnetic moment, can be related to using Ising symmetry of the Hund exchange, as discussed in Ref.~\onlinecite{Sangiovanni}.    

In Fig. \ref{FigJq_gFe}, we show the momentum dependence of the exchange interaction in $\gamma$ (fcc) iron at $\beta=10$~eV$^{-1}$. In contrast to $\alpha$-iron and the previous supercell study of $\gamma$-iron \cite{AKAS2}, but in agreement with earlier DFT analysis \cite{gFe} and the calculation of the momentum dependence of the non-local bubble in $\gamma$-iron \cite{OurGamma}, the momentum dependence of the exchange interaction has maxima at the incommensurate positions along the $\Gamma-X$ and $\Gamma-L$ directions, which corresponds to incommensurate correlations. We obtain the maximal exchange interaction $J_{\bf Q}\simeq 0.04$~eV, {which is substantially smaller than in $\alpha$-iron due to frustration effects \cite{gFe,Gornostyrev}}. For $\gamma$-iron, the results of the three considered averages of the exchange interaction yield somewhat different results. In particular, the leading-order approximation (\ref{ExALMIES}) yields a shallow minimum at the $\Gamma$ point (cf. the results of the supercell approach \cite{AKAS2}), while the the consideration of orbital-summed susceptibilities in Eq. (\ref{JqAvDef}) and the simplified average (\ref{JqAv}) of Eq. (\ref{ExMy}) yields stronger minimum. At the same time, the latter approach shows the local minimum at the $X$ point of the Brillouin zone, while the inflection point at $X$ is present in the other approaches. 

The momentum dependence of the exchange interaction in nickel is shown in Fig.~\ref{Jq1}. One can see that
Eq.~(\ref{JqAvDef}), considering the orbital-summed susceptibilities yields an approximately two times smaller exchange interaction than the average of the orbital-resolved exchange interactions of Eq.~(\ref{ExMy}) or Eq.~(\ref{ExALMIES}). The obtained exchange interactions are well fit by the nearest-neighbor approximation; cf. Ref.~\onlinecite{NNNFe1}. The exchange interaction at the Curie temperature $T_C\simeq 800$~K is estimated as $J_0(T_C)\simeq 1.04 $~eV with the use of the orbital-summed susceptibilities. Although the ``thermodynamic consistency" $J_0(T_C)\approx \chi_{\rm loc}^{-1}(T_C)$ (up to the small subtracted local part of the interaction) of the considered approach with orbital-summed susceptibilities is preserved in this case too, the DMFT $T_C$ is not represented by simple mean-field form for the Heisenberg model ($T_C\neq J_0 \mu^2_{\rm loc}/(3 g^2 \mu_B^2)$), since the temperature dependence of the inverse local magnetic susceptibility $\chi_{\rm loc}(T)$ shows at $T\sim T_C$ the crossover between the two different linear dependencies, see Ref.~\onlinecite{Sangiovanni}. We note that even for the linear temperature dependence of the inverse local spin susceptibility $\chi_{\rm loc}^{-1}=3(g\mu_B)^2 (T+\theta)/\mu_{\rm loc}^2$ for weak itinerant magnets one has to account for the substantial Weiss temperature $\theta\approx \sqrt{2}T_K$ ($=1150$~K for the low-temperature susceptibility of nickel), which is related to the Kondo temperature of screening of local magnetic moments $T_K$, see Refs.~\onlinecite{Sangiovanni,MyComment} and references therein). 

Using the low-temperature magnetic moment $\mu^2_{\rm loc}\simeq 2.04 \mu_B^2$, we find effective spin $p_{\rm loc}=0.37$. The corresponding magnon dispersion at $\beta=10$~eV$^{-1}$ is shown in Fig. \ref{FigD}(b). One can see that the magnetic exchange, extracted from the orbital-summed susceptibilities, agrees qualitatively with the experimental data, although somewhat overestimates the spin stiffness. At low temperatures, we find $J_0\simeq 1.2$~eV from Eq. (\ref{JqAvDef}) with orbital-summed susceptibilities; the corresponding spin-wave stiffness $D=J_0 p_{\rm loc} a^2/12\simeq 0.46$~eV$\cdot$\AA$^2$ {which is also somewhat larger than the experimental value $D\simeq 0.4$~eV$\cdot$\AA$^2$, obtained in Ref. \onlinecite{DNi}}. {The other two considered approaches (Eq. (\ref{ExMy}) or its expansion near the local theory, Eq. (\ref{ExALMIES})) yield a much larger exchange interaction, which does not allow one to achieve even qualitative agreement with the spin-wave stiffness and previous DFT analysis  \cite{NoteNi}. {The strong difference of the exchange interactions obtained within the latter approaches compared to the results from orbital-summed susceptibilities is related to the weaker off-diagonal components of the exchange interactions than in iron. 
In all considered cases we therefore find that the orbital-summed susceptibilities provide more reasonable results.
}
}  

  \begin{figure}[t]
		\center{
		\includegraphics[width=0.9\linewidth]{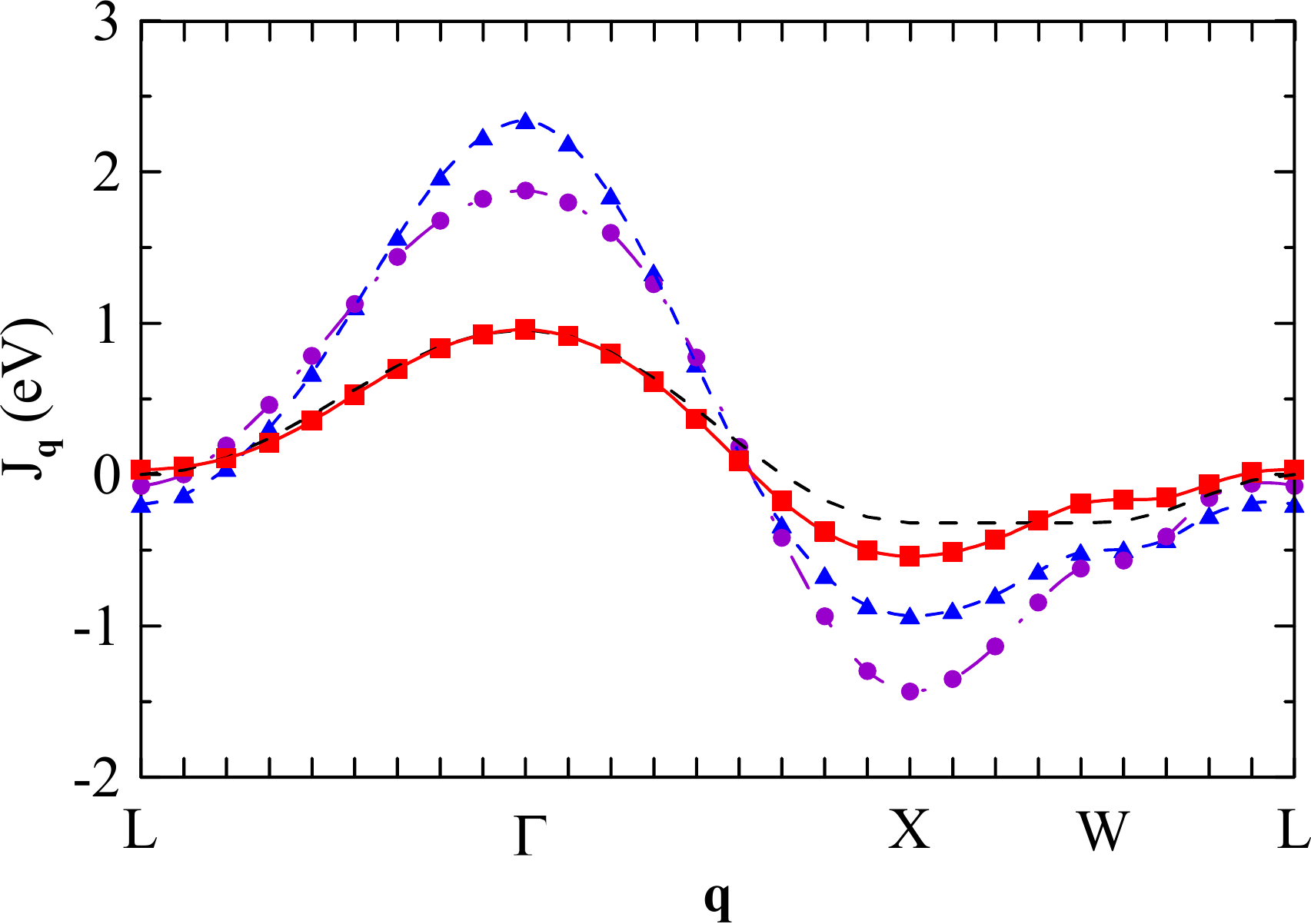}}
		\caption{
		Momentum dependence of the exchange interaction in Ni at $\beta=10$~eV$^{-1}$ along the symmetric directions. The notations are the same as in Fig. \ref{FigJq_aFe}.}
\label{Jq1}
\end{figure}

\section{Conclusion}
In summary, we have obtained momentum dependence of the exchange interaction in the $\alpha$ and $\gamma$ phases of iron, as well as in nickel. 
In particular, in the low-temperature limit of $\alpha$-iron, we obtain from the orbital-summed susceptibilities  $J_0=0.2$~eV in agreement with the supercell DMFT approach. {This approach also allows a correct description of the experimental data for the spin-wave stiffness at $T\sim T_c$. The averaged exchange, obtained from the leading non-local correction, given by Eq. (\ref{ExALMIES}), is somewhat larger, $J_0\simeq 0.26$~eV, and yields larger spin stiffness}. In $\gamma$ iron the exchange interaction is maximal at the incommensurate wave vectors along the $\Gamma-L$ and $\Gamma-X$ direction with the maximal value $J_{\bf Q}\simeq 0.04$~eV which is small due to frustration effects. In Ni, we find $J_0\simeq 1.2$~eV, in agreement with the experimental data for spin-wave stiffness. 
Therefore, the presented method of calculation of exchange interactions yields reasonable results for different substances: strong local-moment magnet $\alpha$-Fe, more itinerant substance with incommensurate correlations $\gamma$-Fe, and weak itinerant magnet nickel. 

The obtained exchange interactions account for correlation effects. 
{Both the Curie temperature 
and spin-wave stiffness 
are better described by the method, considering inverted orbital-summed susceptibilities.}
The developed method also allows one to treat the temperature dependence of exchange interactions. 

Overall, using the DFT+DMFT approach in the paramagnetic phase allows one to obtain reasonable agreement with the experimental data. This method can be used for a wide range of magnetic substances.

{\it Acknowledgements}. Performing the 
DMFT calculations was supported by the Russian Science Foundation
(project 19-12-00012). A.S.B. (in particular, performing the DFT calculations) was supported by the Alexander von Humboldt Foundation. The calculations were performed on the cluster of the Laboratory of Material Computer Design of MIPT and the Uran supercomputer at the IMM UB RAS.

\appendix

\section{Account of the effect of finite frequency box} 

Picking out the contribution of the finite frequency box $B$, we find (cf. Refs.~\onlinecite{My_BS,MyEDMFT})
\begin{eqnarray}
{\hat{\Pi}}_{\bf q}^{}  &=&\sum_{\nu\in B} \hat{\chi}_{{\bf q},\nu} ^{0}\hat{\gamma}_{{\bf q},\nu}^{} +{\hat X}, \label{phi1}
\end{eqnarray}
${\hat X}^{{m}{m}'}=\sum_{\nu _{n}\notin B}{\hat \chi} _{{\rm loc},\nu}^{0,{m}{m}'}$. Considering the contribution of frequencies within and outside the frequency box to the triangular vertex, we obtain \cite{My_BS,MyEDMFT}
\begin{equation}
    {\hat \gamma}_{{\bf q},\nu}^{{m} {m}'}=\sum\limits_{\nu'} \left[\hat{E}-(\hat{\Phi}_{\nu\nu'}^{}-{\tilde{U}}^s) \hat{\chi} ^{0}_{{\bf q},\nu'}\right]_{\nu{m}, \nu'{m}'} ^{-1},\label{gamma1}
\end{equation}
and
\begin{equation}
    {\hat \gamma}_{{\rm loc},\nu}^{{m} {m}'}=\sum\limits_{\nu'} \left[\hat{E}-(\hat{\Phi}_{\nu\nu'}^{}-{\tilde{U}}^s) \hat{\chi} ^{0}_{{\rm loc},\nu'}\right]_{\nu{m}, \nu'{m}'} ^{-1},\label{gammaloc}
\end{equation}
where $\widetilde{U}^{s}=[(\hat{U} ^{s})^{-1}- {\hat X}]^{-1}$. The vertex ${\hat \Phi}$ is evaluated from the local vertex ${\hat F}$ in the spin channel via the Bethe-Salpeter equation (\ref{local_BS}) of the paper
where the inversion is restricted to the frequency box (the effect of frequencies outside the frequency box is accounted by $\hat{X}^{mm'}$ in the Eqs. (\ref{phi1})--(\ref{gammaloc}); see Refs. \onlinecite{My_BS,MyEDMFT}).  

\vspace{-0.6cm}
\section{Relation of the exchange interactions of Eq. (\ref{ExMy}) to the approach of Refs. \onlinecite{ALMIES1,ALMIES2,ALMIES3}} 

In this Appendix, we establish the relation of the result (\ref{ExMy}) to the results of Refs.~\onlinecite{ALMIES1,ALMIES2,ALMIES3}. To this end we expand Eq. (\ref{ExMy}) of the paper to leading order in the non-local part of the Green's function $\widetilde{G}_{\bf k,\nu}=\hat{G}_{{\bf k},\nu}-\hat{G}_{{\rm loc},\nu}$. Representing $\hat{\Pi}_{\bf q}=\hat{\Pi}_{\text{loc}%
}+\delta\hat{\Pi}_{\bf q}$ and considering the diagrammatic form of $\hat{\Pi}_{\bf q},$ in the leading
order we find (cf. Ref.~\onlinecite{Krien})
\begin{equation}
\delta\hat{\Pi}_{\bf q}=\sum_{\nu}\hat{\gamma}^T_{\text{loc},\nu}%
\widetilde{\chi}^0_{{\bf q},\nu}\hat{\gamma}_{\text{loc},\nu},
\end{equation}
where $\widetilde{\chi}^0_{{\bf  q},\nu}=-T\sum_{\mathbf{k}}\widetilde{G}_{{\bf k},\nu}%
\widetilde{G}_{{\bf k+q},\nu}$. 

Expanding Eq. (\ref{ExMy}) in $\delta\hat{\Pi}_{\bf q}$ yields  Eq. (\ref{ExALMIES}) of the paper.
Finally, relating the vertex $\hat{\Lambda}_{\nu}={\hat \Gamma}_{\nu}(2{\hat \chi}_{{\rm loc}})^{-1}$ of Refs.~\onlinecite{ALMIES1,ALMIES2,ALMIES3}, where $\hat{\Gamma}^{mm'}_{\nu}=G_{{\rm loc},\nu m}^{-2}\langle c_{\nu m}^{+}c_{\nu m} S^{z,m'}_{\omega=0}\rangle$, to the vertex $\hat{\gamma}_{{\rm loc},\nu}$, we find
\begin{equation}
\hat{\Lambda}_{\text{loc},\nu}=\hat{\gamma}_{\text{loc},\nu}\hat{\Pi
}_{\text{loc}}^{-1},
\end{equation} 
We then have, in the lowest order in the non-local degrees of freedom,
\begin{equation}
J_{\mathbf{q}}^{\text{l.o.}}=2\sum_{\nu}\hat{\Lambda}^T_{\text{loc},\nu}\widetilde{\chi}^0_{\mathbf{q},\nu}\hat{\Lambda}_{\text{loc},\nu}, \label{Jqlo}%
\end{equation}
which agrees with the result of Refs.~\onlinecite{ALMIES1,ALMIES2,ALMIES3}.





\clearpage
\begin{widetext}
\appendix
\renewcommand\theequation{A\arabic{equation}}
\renewcommand\thefigure{S\arabic{figure}}
\renewcommand\thetable{S\arabic{table}}
\setcounter{equation}{0}
\setcounter{figure}{0}
\setcounter{table}{0}
\setcounter{page}{1}
\section*{Supplemental Material 
\texorpdfstring{\\ \lowercase{to the paper }``E\lowercase{xchange interactions in iron and nickel:} DFT+DMFT \lowercase{study in paramagnetic phase}"}{}}
\vspace{-0.4cm}
\centerline{A. A. Katanin, A. S. Belozerov, A. I. Lichtenstein, M. I. Katsnelson}
\vspace{0.5cm}

\subsection{The momentum dependence of spin susceptibilities  }

Here we present additional results for the momentum dependencies of spin susceptibilities. In Figs. \ref{chiphiqFe} and \ref{chiphiqNi} we show the resulting momentum dependencies of the polarization function ${\Pi}^{}_{\bf q}=\sum_{{ m} { m}'}\hat{\Pi}_{\bf q}^{{ m} { m}'}$ and the non-local susceptibility ${\chi}^{}_{\bf q}=\sum_{{ m} { m}'}\hat{\chi}_{\bf q}^{{ m} { m}'}$ in $\alpha$-Fe and nickel. 
One can see that in agreement with previous results, at the considered temperature the obtained DMFT solution is on the verge of the magnetic phase transition. At the same time, the momentum dependence of ${\Pi}_{\bf q}$ remains smooth and finite.

\begin{figure}[h!]
		\center{
				\vspace{0.3cm}
		\includegraphics[width=0.7\linewidth]{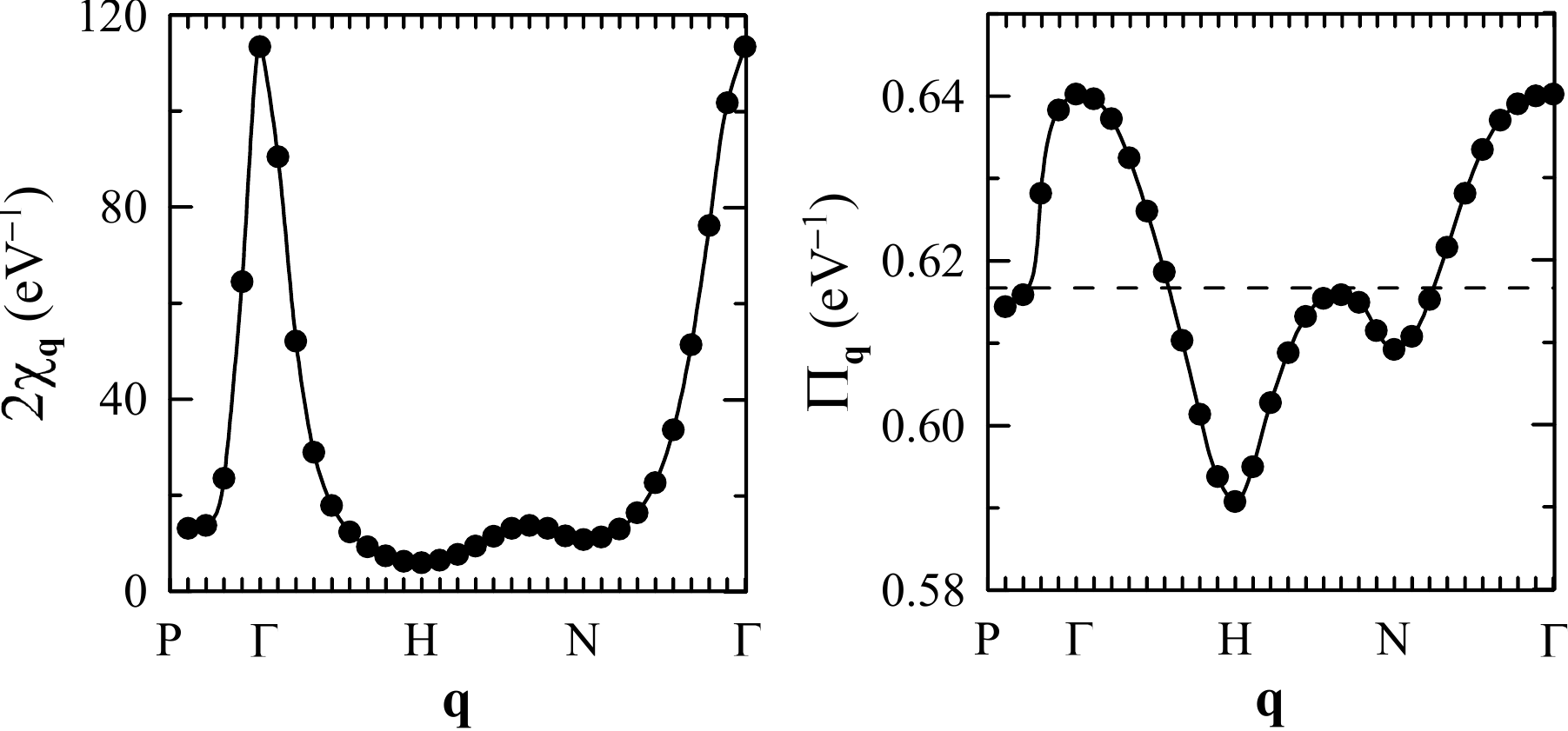}}
		\caption{
		Momentum dependence of the non-uniform susceptibility (left) and irreducible susceptibility (right) in $\alpha$-Fe at $\beta=5$~eV$^{-1}$ along the symmetric directions. The dashed line shows the local value of the irreducible susceptibility.}
\label{chiphiqFe}
\end{figure}

\begin{figure}[h!]
		\center{
				\vspace{0.3cm}
		\includegraphics[width=0.7\linewidth]{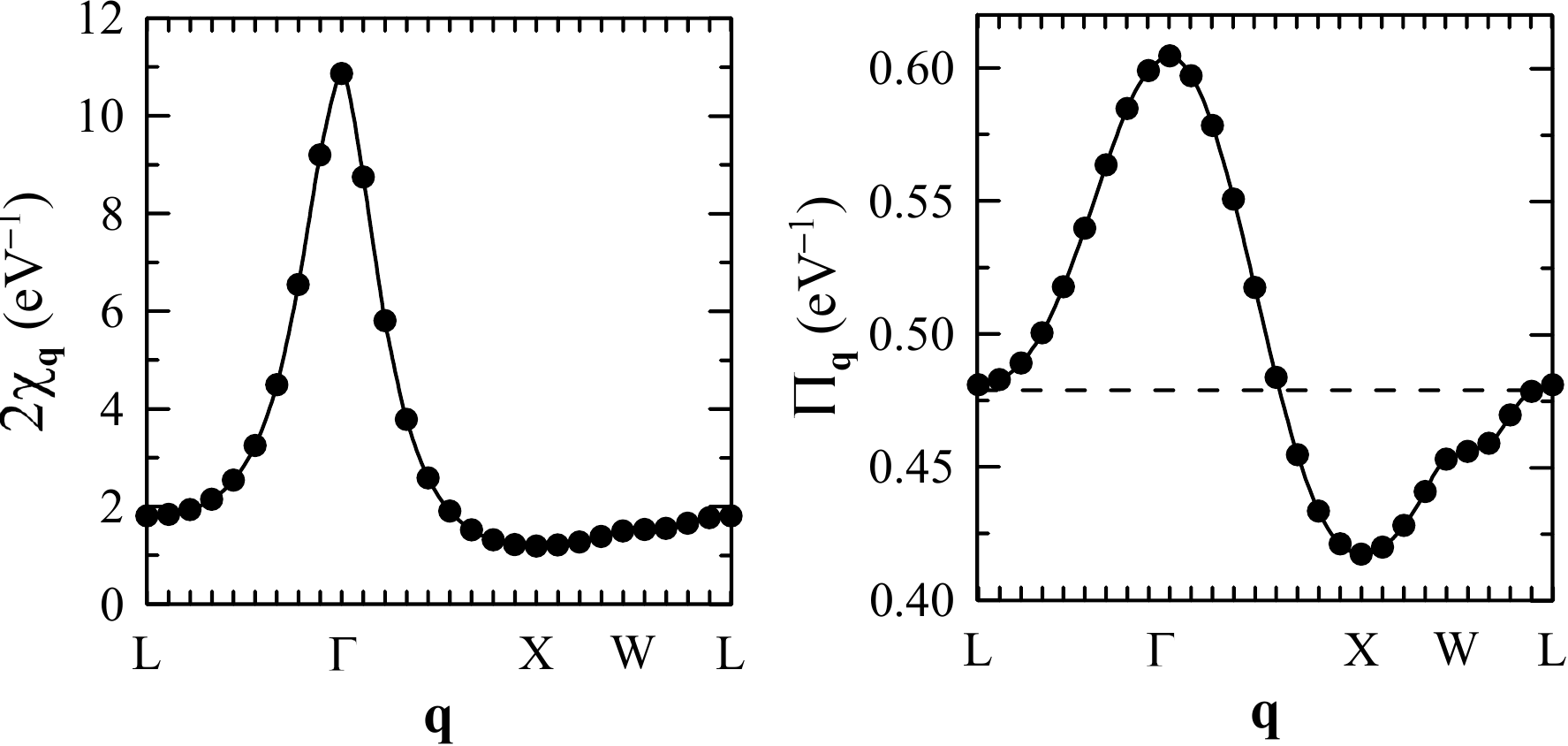}}
		\caption{
		Momentum dependence of the non-uniform susceptibility (left) and irreducible susceptibility (right) in nickel at $\beta=10$~eV$^{-1}$ along the symmetric directions. The dashed line shows the local value of the irreducible susceptibility.}
\label{chiphiqNi}
\end{figure}

\subsection{Temperature- and interaction dependence of magnetic exchange}

In Fig. \ref{FigJq_aFe_S1} we present the temperature dependence of magnetic exchange interactions in $\alpha$-iron and nickel. With decrease of temperature, magnetic exchange interaction $J_0$ increases and the maximum of the exchange interaction at ${\bf q}=0$, considered in the paper, is continuously formed. In $\alpha$-iron at very high temperatures (see, e.g. $\beta=1$~eV$^{-1}$ curve) the sign of the exchange interaction is inverted, which seems to be related with activation of electronic states, which are far from the Fermi level, cf. Ref. \cite{Heine} of the main text. At the same time, for nickel the exchange interaction remains positive at zero momentum.

  \begin{figure}[h!]
		\center{
		\includegraphics[width=0.45\linewidth]{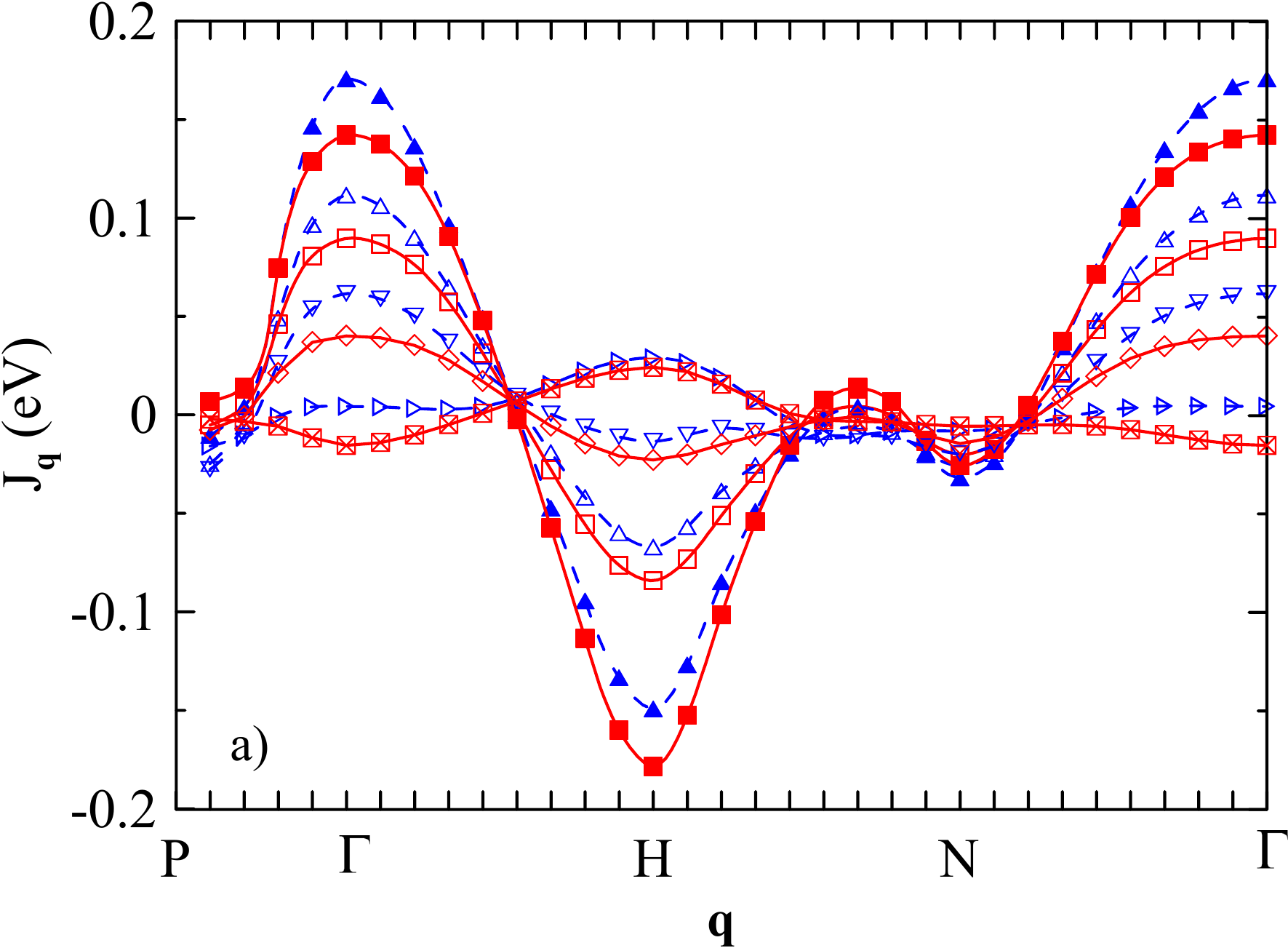}
  \includegraphics[width=0.45\linewidth]{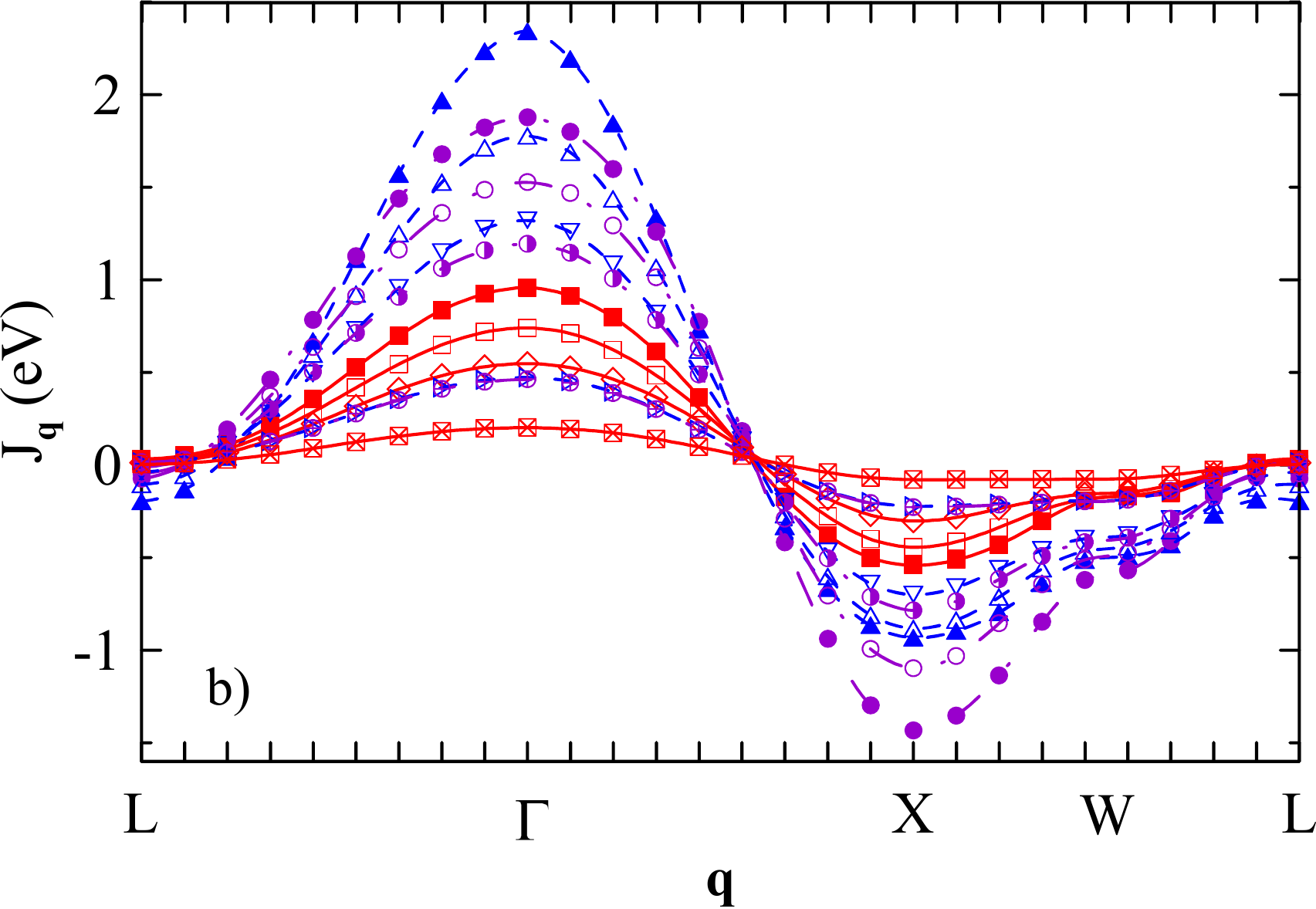}}
		\caption{
		Momentum dependence of exchange interactions along the symmetric directions in (a) $\alpha$-Fe at $\beta=1$~eV$^{-1}$ (crossed squares/right directed triangles), $\beta=2$~eV$^{-1}$ (rhombs/downward directed triangles), $\beta=3$~eV$^{-1}$ (open squares/ open triangles), compared to $\beta=5$~eV$^{-1}$ (filled squares/triangles) and (b) in nickel at $\beta=1$~eV$^{-1}$ (crossed squares/right directed triangles/crossed open circles), $\beta=3$~eV$^{-1}$ (rhombs/downward directed triangles/half filled circles), $\beta=5$~eV$^{-1}$ (open squares/ open triangles), compared to $\beta=10$~eV$^{-1}$ (filled squares/triangles). Blue dashed lines correspond to the leading order approximation (\ref{ExALMIES}) with the average (\ref{JqAv}), violet dashed lines in (b) - to the result from the inverse susceptibilities (\ref{ExMy}) with the average (\ref{JqAv}), red solid lines -- to the result from the orbital summed susceptibilities, Eq. (\ref{JqAvDef}).}
		 \label{FigJq_aFe_S1}
\end{figure}

  \begin{figure}[h!]
		\center{
		\includegraphics[width=0.48\linewidth]{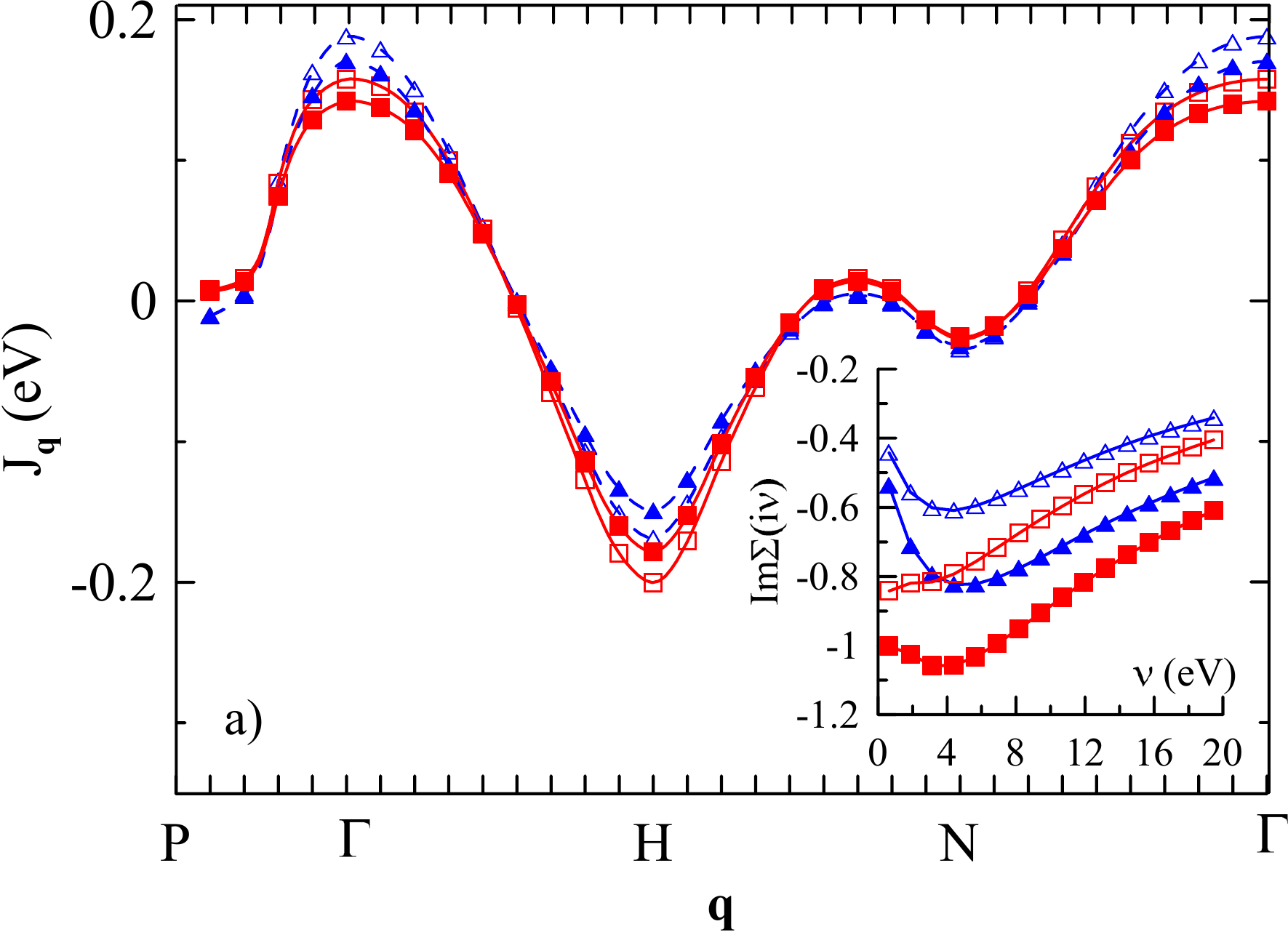}
		\includegraphics[width=0.48\linewidth]{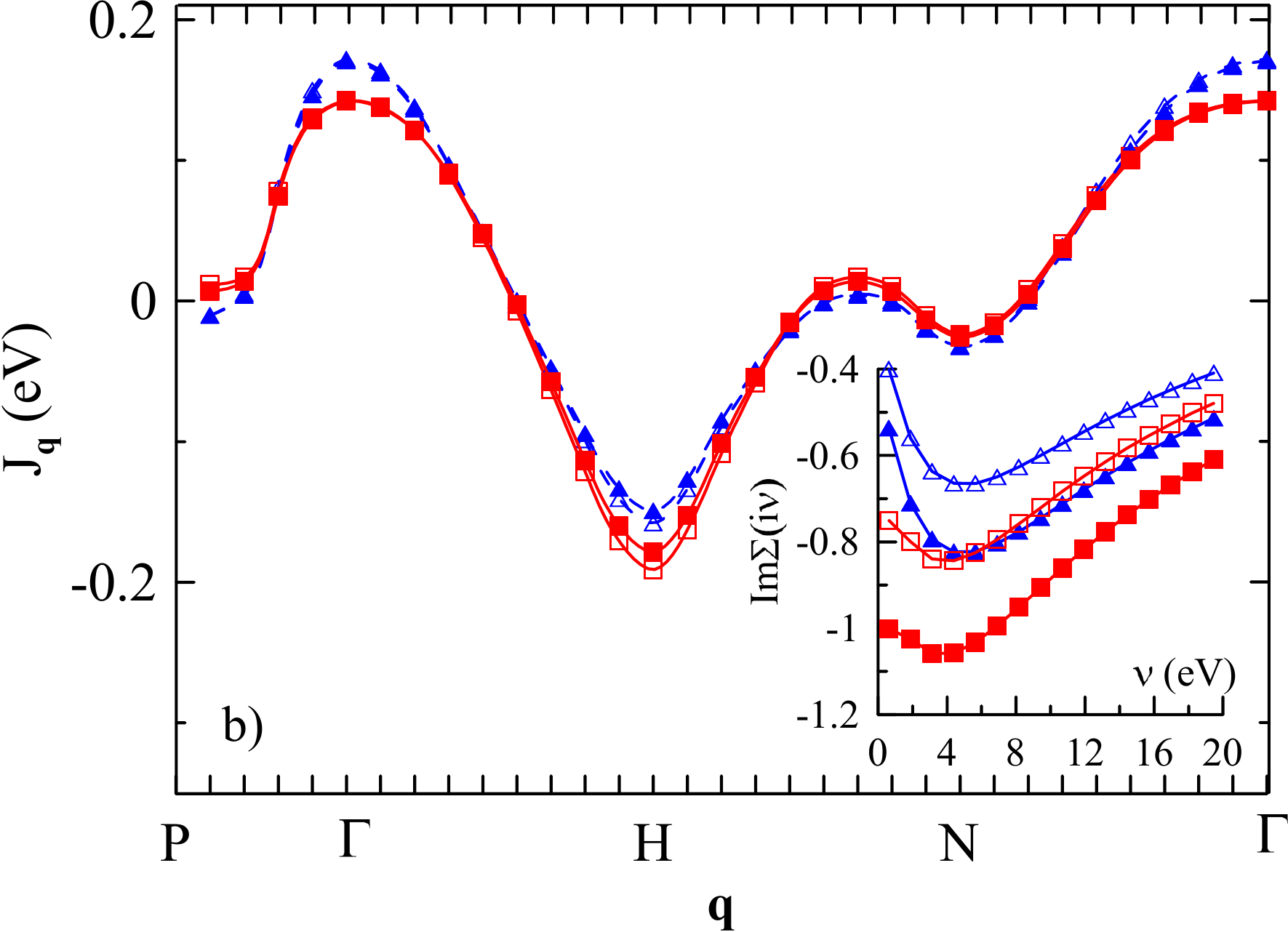}}
		\caption{
		Momentum dependence of exchange interactions in $\alpha$-Fe at $\beta=5$~eV$^{-1}$ along the symmetric directions (main panel) and the corresponding electronic self-energies at the imaginary frequency axis (inset) at (a) $U=3$~eV, $J_{\rm H}=0.9$~eV and (b)  $U=3.5$~eV, $J_{\rm H}=0.8$ ~eV (open squares/triangles) compared to $U=4$~eV, $J_{\rm H}=0.9$~eV (filled squares/triangles). In the main panel blue dashed lines (triangles) correspond to the leading order approximation (\ref{ExALMIES}) with the average (\ref{JqAv}), red solid lines (squares) -- to the result from the orbital summed susceptibilities, Eq. (\ref{JqAvDef}). In the inset open/filled squares correspond to the $e_{g}$ states and open/filled triangles to the $t_{2g}$ states.}
		 \label{FigJq_aFe_S2}
\end{figure}

  \begin{figure}[b]
		\center{
		\includegraphics[width=0.45\linewidth]{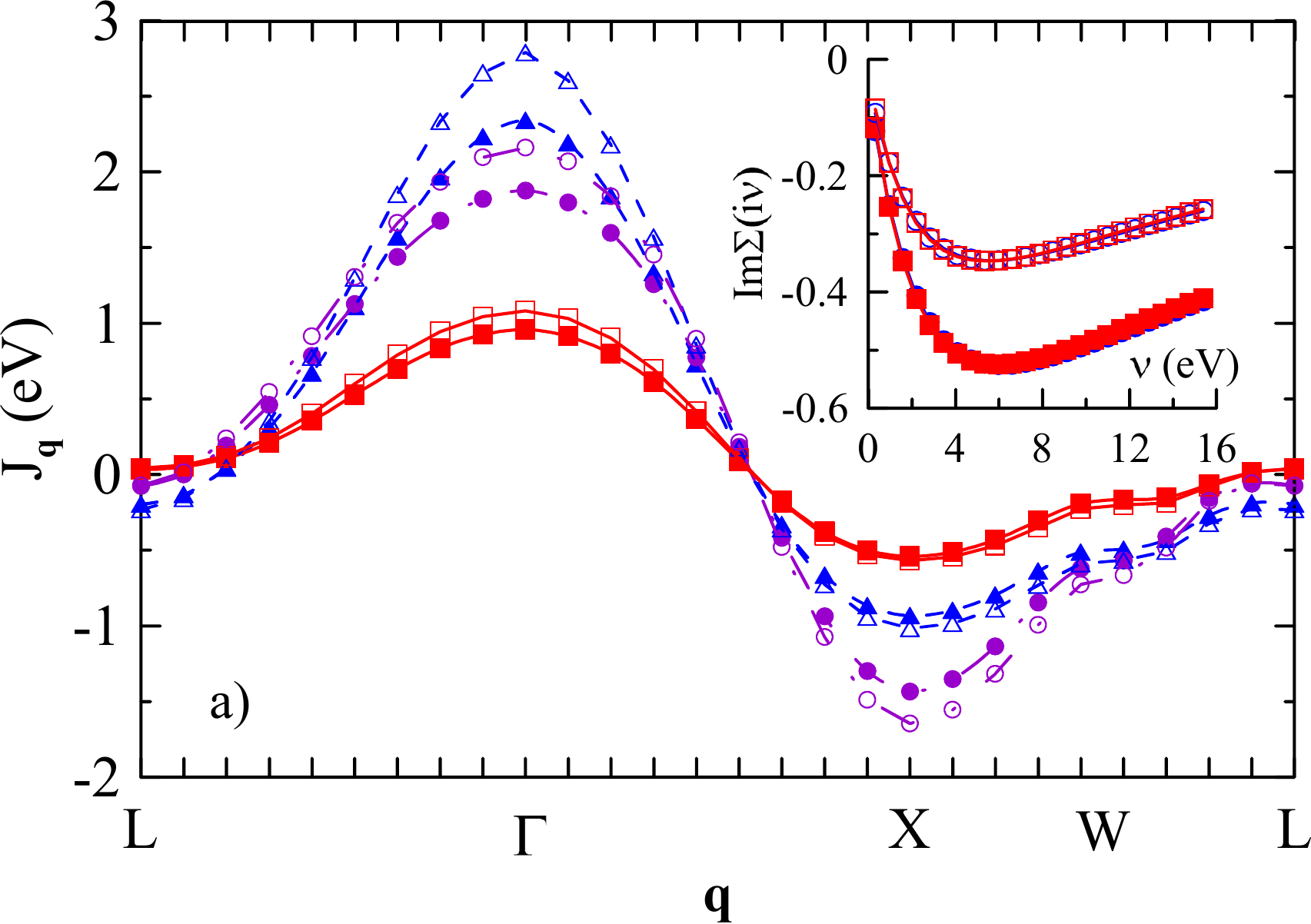}
  		\includegraphics[width=0.45\linewidth]{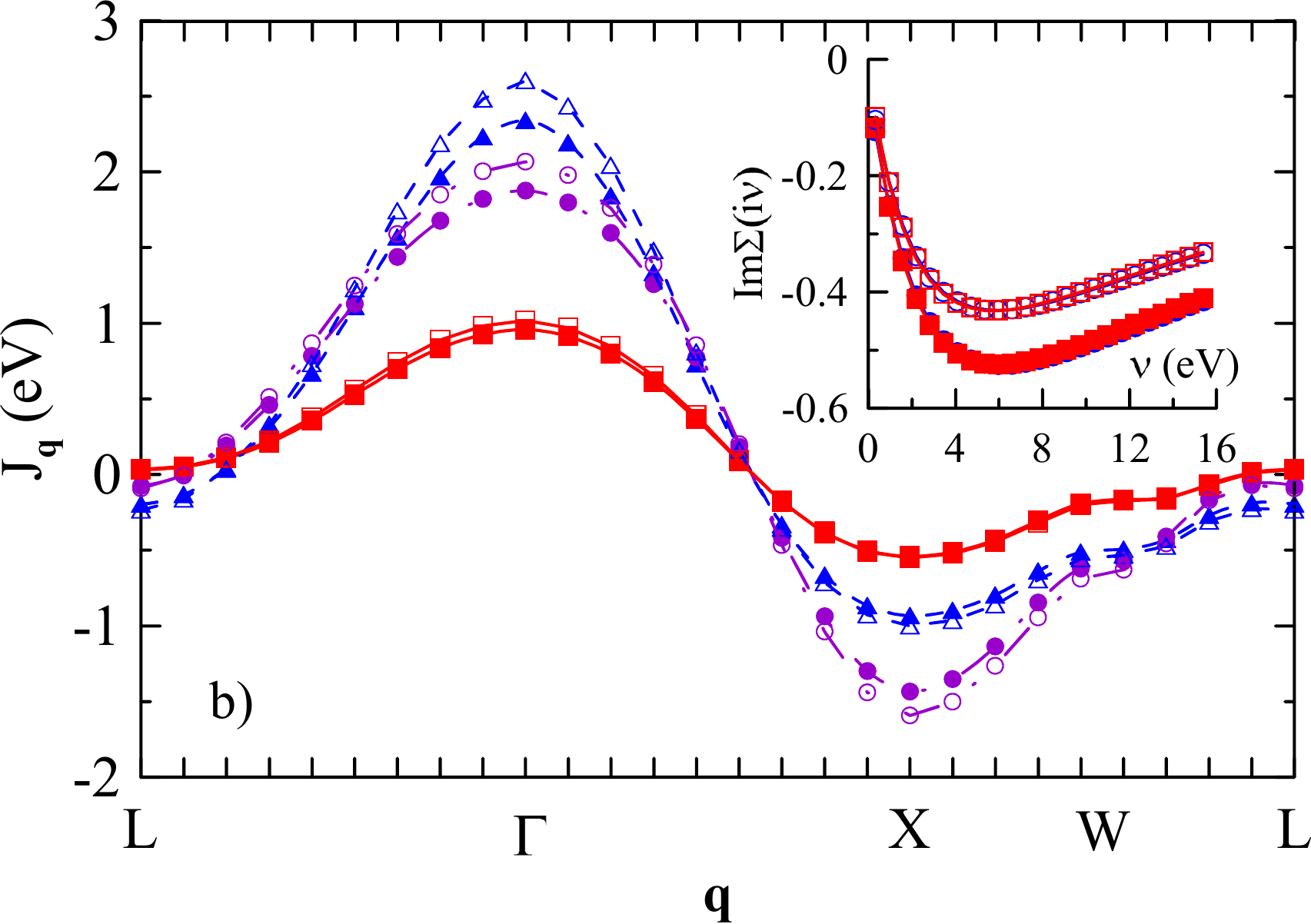}}
		\caption{
The same as Fig. \ref{FigJq_aFe_S2} for nickel at $\beta=10$~eV$^{-1}$. }
		 \label{FigJq_aFe_S3}
\end{figure}

In Fig. \ref{FigJq_aFe_S2} and \ref{FigJq_aFe_S3} we present the interaction dependence of magnetic exchang interaction in iron and nickel. For fixed $J_{\rm H}=0.9$~eV decrease of the interaction to $U=3$~eV (Fig.  \ref{FigJq_aFe_S2}) yields slight increase of magnetic exchange due to increasing role played by Hund exchange interaction $J_{\rm H}$ (which also yields {\it stronger} non-quasiparticle form of the electronic self-energy, see the insets of Fig. \ref{FigJq_aFe_S2}). At the same time, the proportional decrease of both, direct and Hund exchange interactions to $U=3.5$~eV and $J_{\rm H}=0.8$~eV practically does not change the exchange interactions in iron and only slightly increases them in nickel (see Fig. \ref{FigJq_aFe_S3}). As we discuss in the main text, this is due to the compensation of the self-energy and vertex corrections in the considered approach.

\end{widetext} 

\end{document}